\documentclass[reprint,amsmath,amssymb,aps]{revtex4-2}

\usepackage{physics} 
\usepackage{amsfonts}
\usepackage{xfrac}
\usepackage{siunitx}
\usepackage{graphicx}
\usepackage{subcaption}
\usepackage{dcolumn}
\usepackage{bm}
\usepackage{hyperref}
\usepackage{xcolor}

\usepackage{pdfpages}

\makeatletter
\AtBeginDocument{\let\LS@rot\@undefined}
\makeatother

\begin{document}
\title{Understanding the Cavity Born--Oppenheimer Approximation}

\author{Marit R. Fiechter}
\affiliation{Department of Chemistry and Applied Biosciences, ETH Zürich, 8093 Zürich, Switzerland}
\author{Jeremy O. Richardson}
\email{jeremy.richardson@phys.chem.ethz.ch}
\affiliation{Department of Chemistry and Applied Biosciences, ETH Zürich, 8093 Zürich, Switzerland}

\date{\today}

\begin{abstract}
  Experiments have demonstrated that vibrational strong coupling between molecular vibrations and light modes can significantly change molecular properties, such as ground-state reactivity. Theoretical studies towards the origin of this exciting observation can roughly be divided in two categories, with studies based on Hamiltonians that simply couple a molecule to a cavity mode via its ground-state dipole moment 
  on the one hand, and on the other hand \emph{ab initio} calculations that self-consistently include the effect of the cavity mode on the electronic ground state within the cavity Born--Oppenheimer (CBO) approximation; these approaches are not equivalent. The CBO approach is more rigorous, but unfortunately it requires the rewriting of electronic-structure code, and gives little physical insight. In this work, we exploit the relation between the two approaches and demonstrate on a real molecule (hydrogen fluoride) that for realistic coupling strengths, we can recover CBO energies and spectra to high accuracy using only out-of-cavity quantities from standard electronic-structure calculations. In doing so, we discover what the physical effects underlying the CBO results are. Our methodology can aid in incorporating more, possibly important features in models, play a pivotal role in demystifying CBO results and provide a practical and efficient alternative to full CBO calculations.
\end{abstract} 
\maketitle

\section{Introduction}

It is no wonder that the young field of polariton chemistry \cite{garcia2021manipulating,dunkelberger2022vibration} has attracted a lot of attention in recent years; its basis is the surprising observation that one can significantly change molecular behaviour by simply placing the molecules between two mirrors, forming a so-called cavity. It has been known for a while that, when placing the mirrors at a very specific distance from each other, such that the wavelengths that can stand between them match a wavelength associated with an atomic/molecular transition, hybrid light-matter states called polaritons are formed. Excitingly, it has recently been observed that under strong coupling of molecular vibrations to a cavity mode (vibrational strong coupling, or VSC \cite{nagarajan2021chemistry,dunkelberger2022vibration}), one can manipulate ground-state chemistry. Specifically, VSC has been shown to lead to significant acceleration or deceleration of thermal reactions \cite{thomas2016ground,thomas2019tilting,lather2019cavity,vergauwe2019modification,hirai2020modulation,sau2021modifying,ahn2023modification}, and alter product ratios \cite{thomas2019tilting,sau2021modifying}, equilibrium constants \cite{pang2020role,ebbesen2023direct}, solvent polarity \cite{piejko_solvent_2023} and conductivity \cite{fukushima2022inherent,kumar2023extraordinary}. However, in spite of a large body of theoretical work \cite{campos2023swinging,mandal2023theoretical,ruggenthaler2023understanding}, the mechanism by which VSC accomplishes this remains elusive. 

Attempts to understand how VSC can affect molecular properties can roughly be divided into two directions of study. One the one hand, we have rate theories and dynamical simulations based on model systems \cite{li2021cavity,fischer2021ground,mandal2022theory,fischer2022cavity,lindoy2022resonant,wang2022cavity,wang2022chemical,philbin2022chemical,du2023vibropolaritonic,lindoy2023quantum,fiechter2023RPMD} or precomputed \emph{ab initio} potentials or force fields \cite{li2021collective,sun2022suppression,sun2023modification,yu2023manipulating,lieberherr2023vibrational} that coupled to a cavity mode via the ground-state dipole moment; on the other hand, we have fully \emph{ab initio} descriptions of cavity as well as the molecules \cite{flick2017cavity,schafer2022shining,bonini2022ab,schnappinger2023PES,schnappinger2023vibr,sidler2023unraveling}. In the latter, the cavity Born--Oppenheimer (CBO) approximation \cite{flick2017atoms, flick2017cavity,fischer2023beyond} plays an important role; it entails treating photons on the same footing as nuclei (motivated by their frequencies lying in the same range as vibrations), and thereby provides one with the cavity-coupled ground-state potential energy surface (PES). Giving photons the same treatment as nuclei means that the photonic potential energy terms, just like the nuclear potential energy terms, have to be included in electronic structure calculations, and thereby influence the electronic ground-state wavefunction. 

This CBO approach has a few disadvantages, a major one being that the photonic potential energy has to be incorporated within electronic structure code. This means that one cannot just use conventional, highly optimized electronic structure packages, which have been developed by a huge community effort over decades of time; instead, one has to write ones own code. It is not unlikely that in the future, efforts will be made to incorporate the CBO in a standard electronic-structure program, but this would involve overhauling the entire program.
Furthermore, and perhaps even more importantly, the relation of the CBO PES to the out-of-cavity PES is opaque. This makes it difficult to understand exactly what effects the cavity has on the molecule. Additionally, one may rightly wonder whether the way in which the aforementioned model systems are coupled to a cavity mode complies with the CBO point of view.

Inspired by a recent article \cite{fischer2023beyond}, we aim to elucidate the connection between these two directions of study, and in particular clarify which effects the aforementioned model Hamiltonians do not capture. We first review the difference between the CBO approach and simply projecting the Hamiltonian on the out-of-cavity ground state, and show how to improve upon the latter with perturbative corrections. We then move on to study the effect of a cavity mode on a real molecule, hydrogen fluoride, for which CBO-Hartree--Fock results are available \cite{schnappinger2023PES,schnappinger2023vibr}. We show that we can reproduce all CBO results to good accuracy: in some cases, the ground-state projection alone is already sufficient, while that in other cases, one needs to include just one or two perturbative corrections for reaching good agreement with the CBO values. We point out that these perturbative corrections have physical meaning: for example, in the case of harmonic spectra, the leading correction exactly corresponds to accounting for the refractive index of the molecules in the cavity, meaning the frequency shift observed in CBO spectra has a much more mundane origin than suggested in recent literature \cite{bonini2022ab,schnappinger2023vibr,sidler2023unraveling}. We emphasize that interpreting correction terms this way can be crucial in demystifying observations made from CBO results.

\section{Results}
\subsection{Theoretical framework}

We start by writing out the Pauli--Fierz Hamiltonian \cite{mandal2023theoretical,ruggenthaler2023understanding}:
\begin{equation}
	\hat{H}_\text{PF} = \hat{T}_\text{N} + \hat{H}_\text{e} +  \sum_{\lambda,  k}^{2N_\text{c}}\Bigg[\frac{p_{\lambda k}^2}{2} + \frac{\omega_k^2}{2}\Bigg(q_{\lambda k} + \sqrt{\frac{2}{\hbar \omega_k}}\eta_k \hat{\mathbf{e}}_{\lambda} \cdot \hat{\boldsymbol{\mu}}  \Bigg)^2 \Bigg]
\end{equation}
where we have taken the long-wavelength approximation and work in the length gauge. In this expression, $\hat{H}_\text{e}=\hat{T}_\text{e} + \hat{V}_\text{ee} +  \hat{V}_\text{eN} + \hat{V}_\text{NN}$ is the standard electronic Hamiltonian, comprised of the electronic kinetic energy term and all (electron--electron, electron--nuclear, nuclear--nuclear) Coulomb interaction terms, and $\hat{T}_\text{N}$ is the nuclear kinetic energy. The remaining terms describe the $N_\text{c}$ cavity modes and their interaction with molecules, with $\omega_{k}$ the frequency of the $k$th cavity mode (which is $k$ times the fundamental frequency $\omega_\text{c}$), $\hat{\mathbf{e}}_\lambda$ a unit vector indicating its polarization direction, $q_{\lambda k}$ the photonic displacement coordinate corresponding to mode $k$ with polarization $\lambda$, the dipole moment operator
$ \hat{\mu}=-\sum_{i=1}^{N_\text{e}}\hat{r}_i + \sum_{j=1}^{N_\text{N}} Z_j \hat{R}_j $
for a system of $N_\text{e}$ electrons and $N_\text{N}$ nuclei with charge $Z_j$ (setting the elementary charge $e=1$), and the coupling strength for mode $k$ is given by $\eta_k = \sqrt{\hbar/2\omega_k \varepsilon_0 \mathcal{V}} = \eta/\sqrt{k}. $
Here, $\mathcal{V}$ is the quantization volume, and $\varepsilon_0$ the permittivity of vacuum. 

At this stage, the problem can be simplified by separating the electronic degrees of freedom from both the nuclear and photonic degrees of freedom in a similar manner to the standard Born--Oppenheimer approximation \cite{flick2017atoms, flick2017cavity,fischer2023beyond}. We rewrite the Pauli--Fierz Hamiltonian as $\hat{H}_\text{PF}=\hat{T}_\text{N} + \hat{T}_\text{c} + \hat{H}_\text{CBO}$, with $\hat{T}_\text{c}$ the photonic kinetic energy and 
\begin{equation}\label{eq:HCBO}
    \hat{H}_\text{CBO}=\hat{H}_\text{e} + \sum_{\lambda,  k}^{2N_\text{c}}\frac{\omega_k^2}{2}\Big(q_{\lambda k} + \sqrt{\frac{2}{\hbar \omega_k}}\eta_k \hat{\mathbf{e}}_{\lambda} \cdot \hat{\boldsymbol{\mu}}  \Big)^2,
\end{equation}
which, in addition to the standard $\hat{H}_\mathrm{e}$ term also includes photonic potential energy operators. 
The CBO electronic states are the eigenstates of $\hat{H}_\text{CBO}$, which parametrically depend on both the nuclear coordinates $\mathbf{R}$ and photonic displacement coordinates $\mathbf{q}$,
\begin{equation}
	\hat{H}_\text{CBO}\ket{\nu} = V_{\text{CBO},\nu}(\mathbf{R}, \mathbf{q})  \ket{\nu}.
\end{equation}
Solving this equation requires the photonic potential (second term in Eq. \ref{eq:HCBO}) to be implemented within the electronic-structure code. The nuclear and photonic kinetic energy operators $\hat{T}_\text{N}$ and $\hat{T}_\text{c}$ causes derivative couplings to arise between the CBO electronic states; in the CBO approximation, these are neglected. The quality of this approximation is discussed in Ref. \cite{fischer2023beyond}.

As an alternative to the CBO approximation, we can avoid having to adapt electronic-structure code by working in the basis of ``bare" electronic states, that is, eigenstates of $\hat{H}_\text{e}$: 
\begin{equation}
	\hat{H}_\text{e}\ket{n} = V_{\text{BO},n}(\mathbf{R})  \ket{n}.
\end{equation}
Note that these are the electronic states one uses in the standard (cavityless) Born--Oppenheimer approximation. They are only equal to the CBO states at certain photonic configurations, \emph{i.e.} $q_{\lambda k}=-\sqrt{2/\hbar\omega_k}\eta_k \hat{\mathbf{e}}_{\lambda} \cdot \hat{\boldsymbol{\mu}}=0$, making it a type of ``crude adiabatic" basis \cite{fischer2023beyond}.  

In this basis of bare electronic states, we can express the $\hat{H}_\text{CBO}$ as
\begin{equation} \label{eq:HcrudeCBO}
	\begin{aligned}
		H_{nn} = & V_{\text{BO},n}(\mathbf{R}) +  \sum_{\lambda,  k}^{2N_\text{c}}\Bigg[ \frac{\omega_k^2}{2}q_{\lambda k}^2  \\
        + & \sqrt{2/\hbar} \,\omega_k^{3/2}  \eta_k \, q_{\lambda k} \, \bra{n}\hat{{\mu}}_\lambda\ket{n} + \frac{\omega_k}{\hbar} \eta_k^2 \bra{n}\hat{{\mu}}_\lambda ^2\ket{n}\Bigg]   \\
		H_{mn} =&  \sum_{\lambda,  k}^{2N_\text{c}}\Bigg[ \sqrt{2/\hbar} \,\omega_k^{3/2}  \eta_k \, q_{\lambda k} \, \bra{m}\hat{{\mu}}_\lambda\ket{n} \\
        &+ \frac{\omega_k}{\hbar} \eta_k^2 \bra{m}\hat{{\mu}}_\lambda ^2\ket{n}\Bigg]  \text{ for } m\neq n ,
	\end{aligned}
\end{equation}
where we defined $\hat{\mu}_\lambda\equiv\hat{\mathbf{e}}_{\lambda} \cdot \hat{\boldsymbol{\mu}}$ for the sake of brevity.

Importantly, in this crude adiabatic basis, the cavity introduces diabatic couplings $H_{mn}$ between electronic states, meaning that the bare electronic ground state $\ket{0}$ is in general not the true ground state anymore after coupling the cavity \cite{fischer2023beyond}. If all elements $H_{mn}$ are known, it is however straightforward to find the CBO ground state: one just needs to diagonalize the ``crude CBO  (cCBO)" Hamiltonian in Eq. \ref{eq:HcrudeCBO}. 

Unfortunately, to diagonalize this Hamiltonian, one needs to calculate all its elements, which is not an easy task as it requires properties of all excited electronic states. Here, we choose a different path; we aim to perturbatively diagonalize the cCBO Hamiltonian and use this to study how the cavity affects the bare electronic ground state. This makes sense as long as the diabatic couplings are small compared to the energy differences between states, or in other words, as long as the light--matter coupling $\eta$ is small, which one would expect to be the relevant regime for infrared Fabry--Pérot microcavities. The validity of this approximation can be tested by evaluating terms that are higher order in $\eta$ and checking that they become progressively smaller. 

We start by defining our zeroth order Hamiltonian as
\begin{equation}
	\hat{H}^{(0)}=\sum_n H_{nn} \ket{n}\bra{n}
\end{equation}
with $H_{nn}$ the diagonal part of the cCBO Hamiltonian as defined in Eq. \ref{eq:HcrudeCBO} (note that this is a slightly different definition than used in Ref. \cite{fischer2023beyond}; they chose their zeroth-order Hamiltonian to be the cavityless Hamiltonian). 
Our choice of $\hat{H}^{(0)}$ makes the perturbation $\Delta \hat{V} = \hat{H}_\text{CBO} - \hat{H}^{(0)}$ purely off-diagonal,
\begin{equation}
	\Delta \hat{V} = \sum_{m\neq n} H_{mn} \ket{m}\bra{n},
\end{equation}
meaning that the first-order correction to the energy vanishes. The second-order correction to the ground state is 
\begin{equation} \label{eq:PT}
E_0^{(2)} = \sum_{n\neq 0 } \frac{|\bra{0}\Delta\hat{ V }\ket{n} |^2}{E_0 - E_n}.
\end{equation}
Here, $E_n=H_{nn}$ (Eq. \ref{eq:HcrudeCBO}) are the eigenvalues of the zeroth order Hamiltonian, which contain terms proportional to $\eta$ and $\eta^2$.
For the sake of brevity, we report the result here for one cavity mode and polarization only;  the full expression can be found in the SI. Assuming that all excited states are bound states (which is true when using an atomic orbital basis set, as is standard practice in quantum chemistry) so that the transition matrix elements can be chosen to be real, we can remove the modulus signs and write
\begin{equation} \label{eq:PTexpression}
\begin{aligned}
    E_0^{(2)}
        = & \eta^2 q^2  \frac{2\omega_\text{c}^3}{\hbar} \sum_{n\neq 0 } \frac{\,\bra{0}\hat{\mu}\ket{n}^2}{V_{\text{BO},0} - V_{\text{BO},n} }\\ 
         +& \eta^3  q  \sqrt{\frac{8\omega_\mathrm{c}^5}{\hbar^3}} \sum_{n\neq 0 } \frac{\bra{0}\hat{\mu} ^2\ket{n} \bra{0}\hat{\mu}\ket{n}  }{V_{\text{BO},0} - V_{\text{BO},n}}  \\
	    + & \eta^4 \, \frac{\omega_\text{c}^2}{\hbar^2} \sum_{n\neq 0 } \frac{\bra{0}\hat{\mu} ^2\ket{n}^2 }{V_{\text{BO},0} - V_{\text{BO},n} }\\
        - &\eta^3 q^3 \sqrt{\frac{8\omega_\text{c}^9}{\hbar^3}}  \sum_{n\neq 0 } \frac{(\bra{0}\hat{\mu}\ket{0} - \bra{n}\hat{\mu}\ket{n}) \bra{0}\hat{\mu}\ket{n}^2 }{(V_{\text{BO},0} - V_{\text{BO},n})^2 } \\
	    + & O(\eta^4)
    \end{aligned}
\end{equation}
where the $\mathbf{R}$-dependence of $V_{\text{BO},n}(\mathbf{R})$ and $\bra{m}\hat{\mu}\ket{n}$ is implicit, and $\hat{\mu}$ refers to the dipole moment along the chosen polarization. The second $\eta^3$ term in this expression arises from the $\eta$-term in the denominator.

 With this $E_0^{(2)}$ correction, we are sure to include all terms up to order $O(\eta^3)$, since the third-order correction is zero again and the lowest order of $\eta$ in the fourth-order perturbative correction is $\eta^4$. We have explicitly included one $\eta^4$ term here for the sake of completeness, as for stationary points the $\eta^2 q^2$ and $\eta^3 q$ terms become of order $\eta^4$ too, making all three equally important (see discussion below). At this stage, one may already recognize that the $\eta^2$ correction is proportional to the electronic polarizability $\alpha$ \cite{mchale2017molecular,jensen2017introduction}; the significance of this will become clear later. Note that, had we chosen $\hat{H}^{(0)}=\sum_n V_\text{BO,n} \ket{n}\bra{n}$ as in Ref. \cite{fischer2023beyond}, we would have obtained the same results for each order in $\eta$, but it would have required evaluation of the third-order perturbative correction; we have essentially chosen a quicker route to the same result.

Before investigating the importance the perturbative corrections to the ground state, it is interesting to have a closer look at the zeroth order Hamiltonian.  It is very similar to the model Hamiltonian used in many previous studies of cavity-modified chemical reactivity \cite{fischer2023beyond}, with one major difference: the dipole self-energy (DSE) term, $ (\eta_k^2 \omega_k/\hbar) \bra{0} \hat{\mu}  ^2\ket{0}$, is proportional to the expectation value of $\hat{\mu} ^2$, and not to the square of the expectation value of $\hat{{\mu} }$, as pointed out before \cite{li2020origin}.  In fact,
\begin{equation} \label{eq:mu2mu2}
\begin{aligned} 
    	\bra{0} \hat{\mu} ^2 \ket{0} &= \sum_n \bra{0}\hat{\mu}  \ket{n} \bra{n} \hat{\mu}  \ket{0}\\ & = \bra{0} \hat{\mu}  \ket{0}^2 + \sum_{n\neq 0 } \bra{0}\hat{\mu}  \ket{n} \bra{n} \hat{\mu}  \ket{0},
\end{aligned}
\end{equation}
meaning that the difference between the two depends on the transition dipole moments to all electronic excited states. 
While not (yet) available as a callable property in electronic structure packages,  $\hat{\mu}^2$ can easily be calculated directly from transition dipole moments between the Hartree--Fock molecular orbitals using the Slater--Condon rules \cite{schnappinger2023PES}, as also outlined in the Methods section; we anticipate that extension of this calculation of $\hat{{\mu}}^2$ to multi-determinant wavefunctions is straightforward. The fact that the DSE depends on the expectation value of $\hat{{\mu}}^2$ and not on that of $\hat{{\mu}}$ has important qualitative consequences \cite{li2020origin}, for example for the energy and position of minima and transition states, as discussed below.

The discussion above can straightforwardly be generalized for $N$ well-separated molecules. Assuming the molecules to be distinguishable, the many-body ``bare" electronic ground state is just the tensor product of $N$ single-molecule ground states: $\ket{0} = \ket{0}_1 \otimes \ket{0}_2 \otimes ... \otimes \ket{0}_N$. The $N$-molecule Hamiltonian is obtained by replacing $\hat{T}_\mathrm{N}+\hat{H}_\mathrm{e}$ by its sum over all molecules, $\sum_i \hat{T}_{i,\mathrm{N}}+\hat{H}_{i,\mathrm{e}}$, and the single-molecule dipole moment by the sum over dipole moments ($\hat{\mu}=\sum_i \hat{\mu}_i$) \cite{cohen1997photons}. We note that generally, molecules at different positions in the cavity experience a different coupling to the displacement field $\eta_k$ (as the field varies sinusoidally); though straightforward to account for, we will not consider this scenario in this work.

In the above, we have formulated a way of approximating the CBO ground state using information of the molecule outside the cavity only. In the remainder of this work, we investigate how well this cCBO approximation works, with and without perturbative corrections. Specifically, we consider hydrogen fluoride (HF) and compare cCBO potential energies and harmonic frequencies calculated at the Hartree--Fock level of theory to recently published CBO-Hartree--Fock results  \cite{schnappinger2023PES,schnappinger2023vibr}. Specifically, we investigate the following potentials (listed in order of accuracy):
\begin{itemize}
    \item the ``$V_{\langle \hat{\mu}\rangle ^2}$-potential", which is often used in model studies. It is the projection of $\hat{H}_\mathrm{CBO}$ on the bare electronic ground state (Eq. \ref{eq:HcrudeCBO}), but with the $\bra{0}  \hat{{\mu}}^2 \ket{0}$  expectation value replaced by $\bra{0} \hat{{\mu}} \ket{0}^2$.
    \item the ``cCBO potential" $V_\text{cCBO} = H_{00}$ (Eq. \ref{eq:HcrudeCBO}), which like the $V_{\langle \hat{\mu}\rangle ^2}$-potential is the projection of $\hat{H}_\mathrm{CBO}$ on the bare electronic ground state, but now keeping the correct $\bra{0}  \hat{{\mu}}^2 \ket{0}$ expectation value.
    \item the cCBO potential with the $\eta^2$ perturbative correction term, and with both the $\eta^2$ and the $\eta^3$ perturbative correction terms (Eq. \ref{eq:PTexpression}).
    \item the CBO potential (with data from Refs. \cite{schnappinger2023PES,schnappinger2023vibr}); this is our reference. 
\end{itemize}
While performing this comparison, we will not only learn something about the nature of cavity interactions captured by the CBO approximation, but also demonstrate a practical way of calculating the effect of the cavity on a PES, with the possibility of checking the accuracy and systematically improving on the result.

\begin{figure}
    \centering\includegraphics[width=0.35\textwidth]{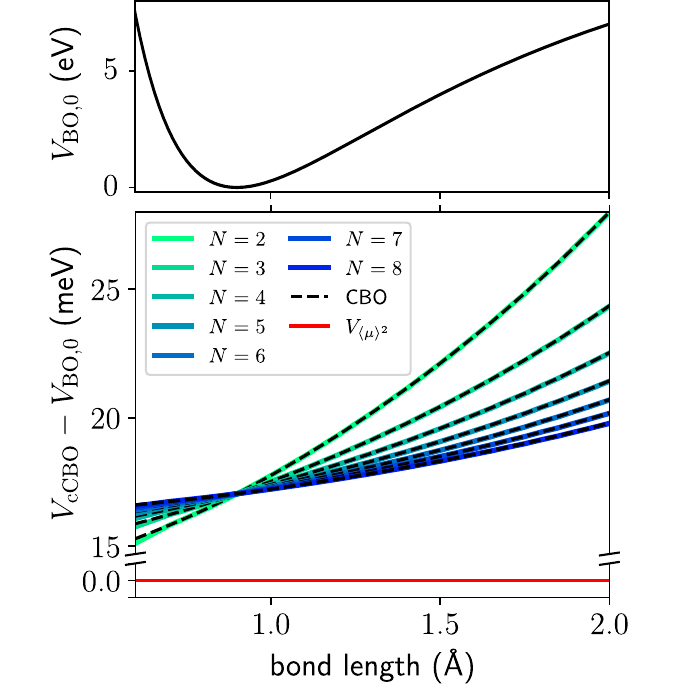}
    \caption{The energy shift caused by the cavity as a function of the bond length $r$ of one HF molecule for stationary values of $q$ (Eq. \ref{eq:qmin}), while the other $N-1$ HF molecules are fixed at their equilibrium bond length ($r_\text{eq} =\SI{0.90}{\angstrom}$ at this level of theory); cCBO (colored lines) compared to CBO reference \cite{schnappinger2023PES} (dashed black lines) and the $V_{\langle \hat{\mu}\rangle ^2}$ potential.}
    \label{fig:bondlength}
\end{figure}

\subsection{Stationary points and their energies}
We from now on consider only a single cavity mode and polarization, and we start out by calculating the position of the stationary points of the cCBO potential.
This has effectively been done before by Li \emph{et al.} \cite{li2020origin}, but for the sake of clarity we reiterate some of their findings here in our notation. Setting $\partial V_\text{cCBO}/\partial q=0$ gives
\begin{equation} \label{eq:qmin}
       q_\text{min} = - \sqrt{2/\hbar} \, \eta \, \omega_\mathrm{c}^{-1/2}  \,  \bra{0} \hat{{\mu}} \ket{0} 
\end{equation}
with $\hat{\mu}=\sum_i \hat{\mu}_i$ the ensemble dipole moment.
Using this, the potential energy along the minimum $q$ is given by
\begin{equation} \label{eq:V_qmin}
    V_\text{cCBO}(\mathbf{R},q_\text{min})=  \sum_{i=1}^{N} \Big[ V_\mathrm{BO,0}(\mathbf{R}_i) + \frac{ \eta^2  \omega}{\hbar} ( \bra{0}  \hat{{\mu}}_i^2  \ket{0} - \bra{0} \hat{{\mu}}_i \ket{0} ^2 ) \Big],
\end{equation}
\emph{i.e.}, the energy of each molecule is shifted by an amount proportional to the variance of $\hat{\mu}$ in the ground state. 

Note that, as the energy shift induced by the cavity depends on the molecular geometry, coupling the cavity can lead to a shift in position of the minimum or transition state. However, this is not a collective effect, as the energy shift of each molecule is independent of the number of molecules in the cavity (or, phrased differently, for a given $\eta$ the total energy in Eq. \ref{eq:V_qmin} scales linearly with $N$), meaning that in practice it is unimportant for studying the modification of chemical properties in microcavities. 

We are now in a position to compare to the CBO-Hartree--Fock results reported in Ref. \cite{schnappinger2023PES}. They performed a scan along the bond length of a HF molecule, which is placed in a cavity along with $N-1$ other HF molecules that are fixed at their equilibrium bond length. All HF molecules are oriented along the polarization direction, and when in increasing the number of molecules in the ensemble, the coupling strength is rescaled by
\begin{equation} \label{eq:rescaling_eta}
    \eta_N=\eta_1/\sqrt{N},
\end{equation}
so as to keep the Rabi splitting approximately constant. The reported calculations were carried out with a cavity frequency equal to the harmonic vibrational frequency of HF, $\omega_\text{c}=\omega_\text{m}=\SI{4467}{\per\centi\meter}$, and at an electric vacuum field strength of $\epsilon_\mathrm{c}=1.0$ V/nm, corresponding to $\eta_1=\epsilon_\mathrm{c}/\omega_\mathrm{c} = 0.0955$ a.u..
Note that the rescaling of $\eta$ keeps this concentration constant when increasing the number of molecules. Further details of the \emph{ab initio} calculations are reported in the Methods section. A plot of the CBO results from Ref. \cite{schnappinger2023PES} compared to our cCBO results is shown in Fig. \ref{fig:bondlength}. For comparison, we have also added the result of the often-used potential $V_{\langle \hat{\mu}\rangle ^2}$, where the $\bra{0}  \hat{{\mu}}_i^2 \ket{0}$  expectation value is replaced by $\bra{0} \hat{{\mu}}_i \ket{0}^2$.

Already from Eq. \ref{eq:V_qmin} it is clear that the $V_{\langle \hat{\mu}\rangle ^2}$ potential captures none of the energy shift along $q_\text{min}$; along stationary $q$ it is equal to $V_\mathrm{BO,0}$ . The cCBO potential however is in very good agreement with the CBO reference results. That it works so well can be rationalized by considering the magnitude of the perturbative corrections (Eq. \ref{eq:PTexpression}) in this case; as $q_\text{min}\propto \eta$, the first three terms all become of order $\eta^4$, and the last of order $\eta^6$, suppressing their importance and rendering the simple cCBO potential very close to the right result already.

\subsection{Potential energy along the photonic displacement coordinate $q_\text{c}$}
We have so far learnt the that cCBO approximation provides a very good estimate of the CBO potential at $q=q_\text{min}$, as the perturbative corrections only come in at $O(\eta^4)$ in that case. An obvious next challenge would then be to move away from $q=q_\text{min}$, and look at the potential as a function of the displacement coordinate $q_\text{c}$ instead; in this case, the leading order correction will be of $O(\eta^2)$.

\begin{figure*}
\includegraphics[width=\textwidth]{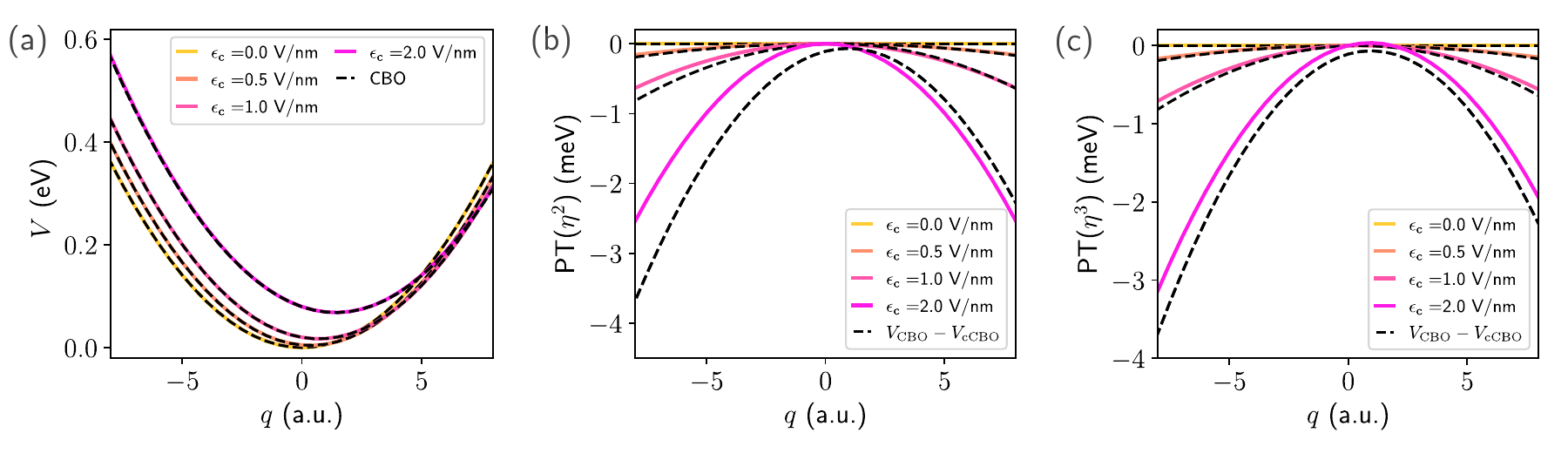}
    \caption{Scan of the potential energy of a single HF molecule along $q$ at fixed $r=r_\text{eq}$ and for $\omega_\mathrm{c}=\SI{4467}{\per\centi\meter}$, without cavity coupling and for 3 different coupling strengths. a) cCBO potential, compared to the CBO results of Ref. \cite{schnappinger2023PES} b) Error in the cCBO potential with respect to the CBO reference, compared to the  perturbative correction of order $\eta^2$ to the cCBO potential. c) as in b), but now additionally including both $\eta^3$ corrections to the cCBO potential. 
    Note that $\epsilon_\mathrm{c}$ is related to $\eta$ by $\eta=\epsilon_\mathrm{c}/\omega_\mathrm{c}$.  }\label{fig:qscan}
\end{figure*}

In Fig. \ref{fig:qscan}a, we show our cCBO potential calculated at the Hartree--Fock level of theory as a function of $q$, and compare it to CBO-Hartree--Fock results \cite{schnappinger2023PES} for one HF molecule fixed at its equilibrium bond length for a range of coupling strengths. We have plotted the results for values of $q$ between $-8$ and $8$ a.u.; this is a sensible range, as 
the equality $\tfrac{1}{2}\omega_\text{c}^2 q^2 = \tfrac{1}{2}\hbar\omega_\text{c}$ (\emph{i.e.} the potential being equal to the zero-point energy) is fulfilled for $q \approx 7.0$ a.u. For the sake of clarity, we have chosen not to display the results for the $V_{\langle\hat{\mu} \rangle^2}$--potential: apart from a constant shift (as discussed previously), the potential energy curves along $q$ are identical to the cCBO curves.

Although the agreement between cCBO and CBO potentials seems quite good to begin with, there are small deviations on the meV scale, especially for larger $q_\text{c}$ (Fig. \ref{fig:qscan}b). To correct for this, we calculate the $\eta^2$ perturbative correction term (``PT($\eta^2$)", first line in Eq. \ref{eq:PTexpression}). This requires evaluation of the transition dipole moments and excitation energies of all excited states. This may seems a daunting task at first, but is dramatically simplified when realizing that from the derivation of perturbation theory it is clear that the required states are excited states of the Hartree--Fock Hamiltonian, and can thus simply be generated from the Hartree--Fock molecular orbitals. 
Additionally, since $\hat{\mu}$ is a one-electron operator, only singly-excited determinants contribute to the sum. This makes evaluation of PT($\eta^2$) rather straightforward. Going on to the $\eta^3$ perturbative correction (``PT($\eta^3$)", second and fourth line in Eq. \ref{eq:PTexpression}) does not require more computational effort, and is described in more detail in the Methods section. 

The perturbative corrections, as well as the deviation between cCBO and CBO potentials that they are supposed to make up for, are plotted in Figs. \ref{fig:qscan}b and \ref{fig:qscan}c. As can be seen in Fig. \ref{fig:qscan}b, the PT($\eta^2$) correction already recovers most of the effects neglected by the cCBO approximation. However, for positive $q_\text{c}$ it slightly overpredicts the correction needed, while for negative $q_\text{c}$ it slightly underpredicts it. Adding the next order correction, PT($\eta^3$) correction, nicely improves upon this, as shown in Fig. \ref{fig:qscan}c.

\subsection{Harmonic spectra}
We now move on to the calculation of harmonic spectra. Rather than the absolute energy, this is an accurate test for the curvature at the minimum geometry. (Note however that we do not reoptimize the HF bond length in the cavity, as this minute effect hardly affects the results and was also neglected in the CBO reference results \cite{schnappinger2023vibr}.) The Hessian of the cCBO Hamiltonian at equilibrium geometry (using Eq. \ref{eq:qmin}) is given in the Methods section. Except for the molecular normal-mode frequencies, cavity frequencies, and off-diagonal terms coupling the cavity to the molecules, it also contains terms coupling the molecules directly to each other (captured by the matrix $\mathbf{G}$) and terms causing a slight shift in the molecular frequency (within the matrix $\mathbf{A}$); these are proportional to $\eta^2$ and 
arise from the $\bra{0}\hat{\mu}^2\ket{0}$ (dipole-self-energy) term. We denote the shifted molecular frequency as $\omega_\mathrm{m}'$. 

\begin{figure*}
    \centering\includegraphics[width=0.95\textwidth]{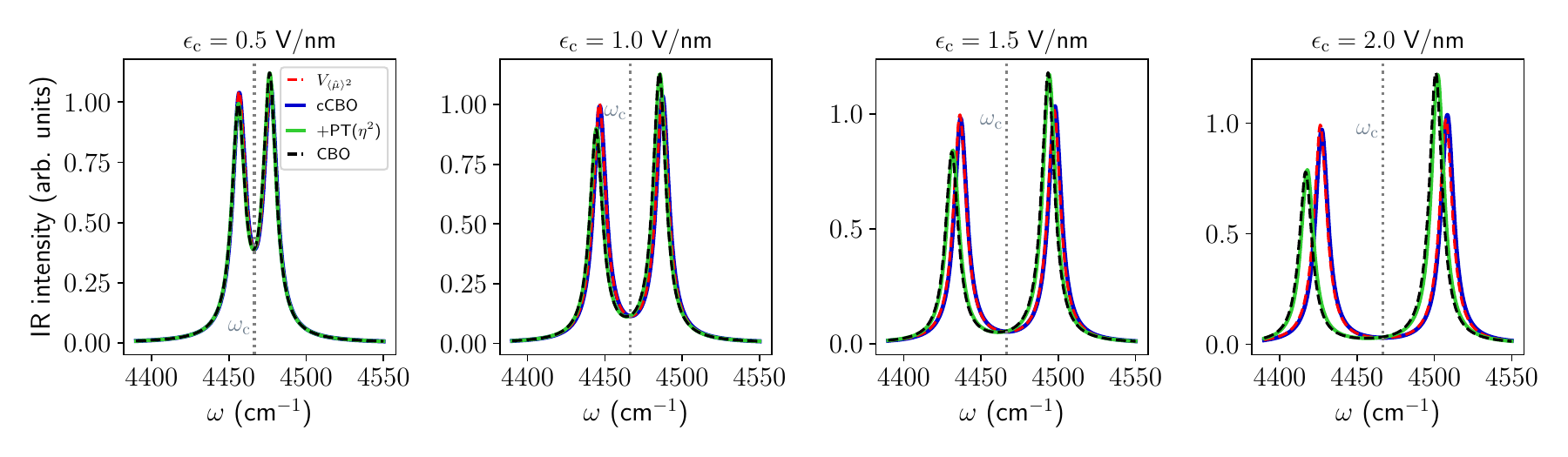}
    \caption{Harmonic spectra for a single HF molecule for a range of coupling strengths; comparing CBO results \cite{schnappinger2023vibr} to cCBO, $V_{\langle\hat{\mu}\rangle^2}$ and cCBO combined with the $\eta^2$ perturbative correction (+PT($\eta^2$)). The cavity frequency was tuned to match the HF harmonic frequency, $\omega_\text{c}=\omega_\text{m}=\SI{4467}{\per\centi\meter}$. }
    \label{fig:spec_1mol}
\end{figure*}

We start by comparing the cCBO spectra for a single HF molecule to the CBO-Hartree--Fock reference results for a range of coupling strengths in Fig. \ref{fig:spec_1mol}. Without perturbative corrections, the cCBO approximation does not improve upon the $V_{\langle\hat{\mu}\rangle^2}$ potential, and it does not reproduce the CBO result well, especially for larger coupling strengths. In particular, both CBO peaks redshift with increasing coupling strength, and the lower polariton peak decreases in intensity; the cCBO peaks on the other hand remain centered around the cavity and HF harmonic frequencies, $\omega_\text{c}=\omega_\text{m}=\SI{4467}{\per\centi\meter}$, and have nearly equal intensities.

To improve, we include the PT($\eta^2$) correction to the cCBO Hamiltonian; the terms it adds to the Hessian are again given explicitly in the Methods section. Interestingly, only the correction to the cavity--cavity block of the Hessian is of $O(\eta^2)$:
\begin{equation}
    \begin{aligned}
        C^\text{PT} 
        &=- \frac{2\omega_\mathrm{c}^3\eta^2}{\hbar}N \alpha_1
    \end{aligned}
\end{equation}
where we have made use of the fact that for frequencies small compared to electronic excitation frequencies, the single-molecule electronic polarizability at equilibrium geometry is defined as $\alpha_1=-2\sum_{n\neq 0 } \,\bra{0}\hat{\mu}\ket{n}^2/(V_0(\mathbf{R}_\text{eq}) - V_{\text{BO},n}(\mathbf{R}_\text{eq})) $ \cite{jensen2017introduction,mchale2017molecular}. The other contributions to the Hessian arising from the PT($\eta^2$) correction have amassed one or two more factors of $\eta$ and are therefore suppressed. The correction effectively \emph{shifts the cavity frequency} \cite{galego2019cavity}, as we will explicitly show below; unlike the slight shift in molecular frequency introduced before, this does not just have a minor effect. In fact, this cavity frequency shift is the origin of the red shift and peak asymmetry observed in the CBO spectra:
as can be seen from Fig. \ref{fig:spec_1mol}, inclusion of the PT($\eta^2$) modifies the cCBO spectrum significantly as to nearly perfectly match the CBO results. We found that further correcting the Hessian with PT($\eta^3$) terms and correcting $q_\text{min}$ by adding the PT($\eta^2$) term to the Hamiltonian before minimizing have a negligible effect (not shown in the Figure). 

\begin{figure}
    \centering\includegraphics[width=0.4\textwidth]{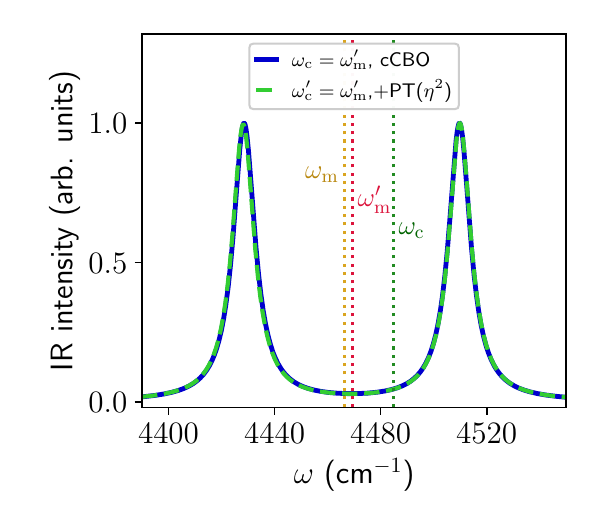}
    \caption{Harmonic spectra for a single HF molecule, for shifted cavity frequencies at a coupling strength of $\epsilon_\mathrm{c}=2.0 $ V/nm, which amounts to a refractive index of $n=1.003962$. \emph{Blue}: cCBO spectrum for $\omega_\mathrm{c}=\omega_\mathrm{m}'=4470$ cm$^{-1}$. 
    \emph{Green}: cCBO+PT($\eta^2$) spectrum,  now while accounting for the refractive index (Eq. \ref{eq:refindex}) when tuning the cavity frequency to resonance, \emph{i.e.} for $\omega_\mathrm{c}'=\omega_\mathrm{m}'$. The corresponding frequency in vacuum in this case is $\omega_\mathrm{c}$ = 4485 cm$^{-1}$.} 
    \label{fig:symmetricspec}
\end{figure}

We now more deeply investigate the cavity frequency shift. Using the definition of $\eta$ in terms of cavity volume, we can express the shifted frequency as
\begin{equation}\label{eq:refindex}
\begin{aligned}
      \omega_\text{c}'^2 &= C+C^\text{PT}\\
      &= \omega_\text{c}^2\Big(1-\frac{2\omega_\text{c}\eta^2}{\hbar} N\alpha_1\Big)\\
      &= \omega_\text{c}^2\Big(1-\frac{N\alpha_1}{\varepsilon_0 \mathcal{V}}\Big).\\
      &= \omega_\text{c}^2/n^2   \hspace{2cm} \Big(\frac{N\alpha_1}{\varepsilon_0 \mathcal{V}} \ll 1\Big)
\end{aligned}
\end{equation}
where we introduced the refractive index $n$, which is given by $n^2=1+N\alpha_1/\varepsilon_0 \mathcal{V}$ in the case of a dilute gas of non-interacting molecules for which the polarizability is purely electronic \cite{mchale2017molecular} (in line with our assumptions; note that our HF molecules are not allowed to rotate, meaning their permanent dipoles do not add to the polarizability). This has a clear physical interpretation: when keeping the cavity length fixed while adding molecules to the cavity, the cavity frequency decreases depending on the index of refraction, so that $\omega_\text{c}'=\omega_\text{c}/n$ \cite{hecht2013optics}. This means that to tune the cavity to resonance, one has to decrease the cavity length or equivalently, increase the vacuum value $\omega_\text{c}$, in order to satisfy $\omega_\text{c}' = \omega_\text{m}' \approx \omega_\text{m}$ (where $\omega_\text{m}'$ includes the aforementioned minor shift arising from the dipole-self-energy term). Doing so results in a pair of equally intense peaks in the IR spectrum, split symmetrically around $\omega_\text{m}'$, as shown in Fig. \ref{fig:symmetricspec}. In fact, it is actually straightforward to see that retuning the cavity to resonance (and only taking $C^\text{PT}$ into account) yields a spectrum that is \emph{equivalent} to the cCBO spectrum tuned to $\omega_\mathrm{c}=\omega_\mathrm{m}'$ (\emph{c.f.} Fig. 4, where, although we did take the entire PT($\eta^2$) correction into account, the differences between both approaches are negligible). In other words, instead of going through the effort of calculating the refractive index and recalibrating the cavity frequency accordingly before doing a CBO calculation, one could just as well perform a cCBO calculation at the vacuum cavity frequency that matches the molecular vibrational frequency, and obtain nearly exactly the same spectrum.

We finally discuss the effects of adding more molecules to the cavity. We note that the $N$-molecule Hessian (explicitly given in the Methods section) is highly structured. For any number of molecules, we can block diagonalize the Hessian, so that all but one $n_\text{modes}\times n_\text{modes}$ molecular block (which is a symmetric linear combination of all molecular blocks) decouple from the cavity mode. The resulting $(n_\text{modes}+1 \times n_\text{modes}+1 )$ matrix is cheap to diagonalize, and its entries are all single-molecule quantities, meaning we never have to perform an \emph{ab initio} simulation of more than one molecule. This is in contrast to the work done in Ref. \cite{schnappinger2023vibr}, in which all 4 molecules are explicitly simulated, and to Ref. \cite{bonini2022ab}, which requires 2-molecule simulations to construct a good approximation to the $N$-molecule Hessian.

\begin{figure*}
    \centering\includegraphics[width=0.95\textwidth]{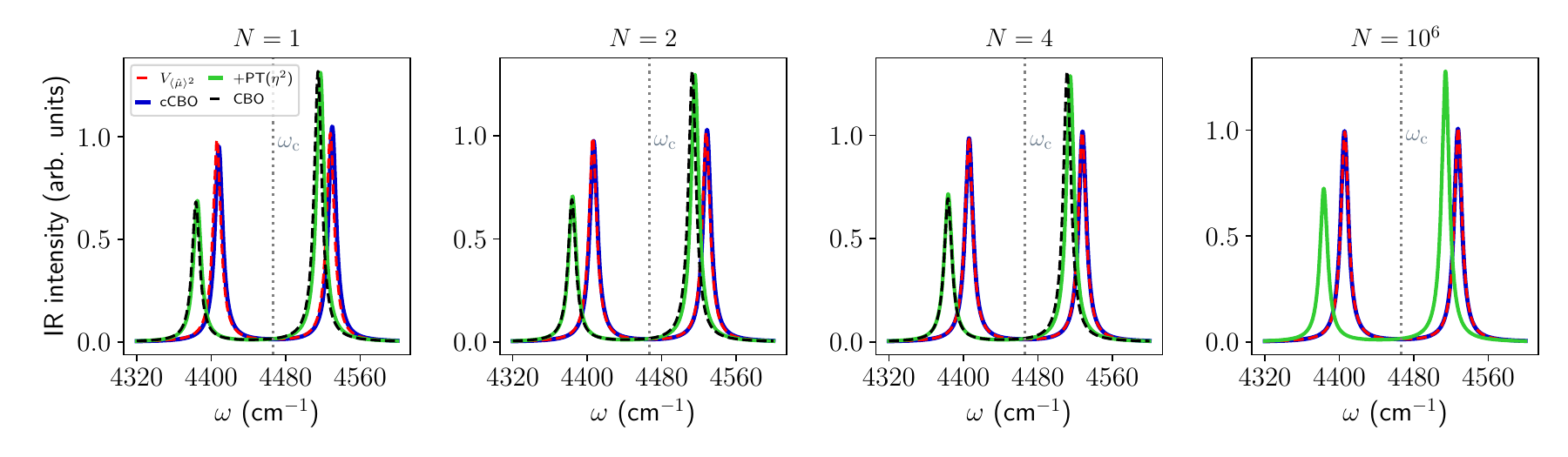}
    \caption{Harmonic spectra for $N=$ 1, 2, 4 and $10^6$ molecules, for a coupling strength corresponding to $\epsilon_\mathrm{c}=$ 3.0 V/nm (rescaled using Eq. \ref{eq:rescaling_eta}), and $\omega_\mathrm{c}=\omega_\mathrm{m}$. }
    \label{fig:Nmolspec}
\end{figure*}

Fig. \ref{fig:Nmolspec} shows harmonic spectra for increasing numbers of molecules for rescaled coupling strengths (Eq. \ref{eq:rescaling_eta}), which indeed results in the Rabi splitting being approximately independent of the number of molecules.
As mentioned before, this rescaling keeps the concentration $N/\mathcal{V}$ and therefore the refractive index constant when increasing the number of molecules; this is in line with the red shift and peak asymmetry of the CBO and cCBO+PT($\eta^2$) results not changing much with increasing $N$. The change with $N$ in both the cCBO and the $\langle\hat{\mu}\rangle^2$ results is also very small; the most noteworthy feature is that both the cCBO and $\langle\hat{\mu}\rangle^2$ peak intensities become more symmetric, so the difference between these two approaches becomes negligible. A detailed comparison of the changes in spectrum with the number of molecules for each method is shown in the SI.

In conclusion, we have investigated the connection between the CBO ground state and out-of-cavity electronic ground state (cCBO approximation), and performed Hartree--Fock calculations on a hydrogen fluoride molecule to explicitly compare to CBO-Hartree--Fock results \cite{schnappinger2023PES,schnappinger2023vibr}; we have been able to reproduce all CBO results using out-of-cavity quantities only.

To start, we have shown that the cCBO energy provides a very good estimate to the CBO energy if one is interested in the potential along the minimum of the displacement coordinate $q_\text{c}$, while in contrast, the often-used $V_{\langle\hat{\mu}\rangle^2}$ potential (where $\langle\hat{\mu}^2\rangle$ is replaced by $\langle\hat{\mu}\rangle^2$) captures none of this. This may not be detrimental for all studies based on this Hamiltonian; the energy shift predicted by cCBO does not scale collectively and is therefore not likely to play a substantial role in for example chemical reactivity. 

The CBO potential along $q_\text{c}$ (at fixed HF bond length) is also well reproduced by its cCBO counterpart. On top of that, we have shown that we can systematically improve upon this by adding perturbative corrections, which are straightforward to evaluate from the output of a Hartree--Fock ground state calculation.

In reproducing the CBO harmonic spectrum of HF, the leading perturbative correction plays a vital role; it shifts the cavity frequency according to the refractive index of the molecules in the cavity. This explains the asymmetry of upper and lower polariton intensities in recent CBO literature \cite{schnappinger2023vibr,sidler2023unraveling,bonini2022ab}: in these studies the \emph{vacuum} cavity frequency was chosen on resonance with a vibrational frequency, meaning that the refractive-index-corrected cavity frequency was off-resonance. To make up for this, the cavity has to be shortened by an amount dependent on the refractive index, corresponding to an increase in vacuum cavity frequency. However, this yields the same spectrum as a much less cumbersome cCBO calculation, in which the refractive index does not need to be accounted for. It is nonetheless interesting to note that in the CBO approximation, information about the refractive index is already contained within the Hamiltonian.

The \emph{ab initio} results in this work were all obtained at the Hartree--Fock level of theory, which does not account for correlation effects between electrons. However, we see no reason to suspect that going to a higher level of electronic structure theory and thereby adding in electron correlation 
would qualitatively alter our findings. 

On the other hand, studying a different molecule at similar concentrations would likely affect the importance of perturbative terms. All perturbative terms are composed of sums over excited states, with each contribution weighted by the inverse of the excitation energy (or higher powers thereof). HF, being a small molecule, has large excitation energies, and consequently \emph{e.g.} a rather small polarizability. Larger molecules tend to have smaller excitation energies and are therefore more polarizable \cite{mchale2017molecular}; by the same reasoning the other perturbative terms are also likely to be larger in magnitude, meaning that one would like need to include more of the terms to converge.

Furthermore, increasing the coupling strength (or equivalently, increasing the concentration by decreasing the volume $\mathcal{V}$) will naturally also increase the importance of perturbative corrections. Nonetheless, we note that the largest coupling strength we have reported results for, $\epsilon_\mathrm{c}=3.0$ V/nm, corresponds to a HF concentration of \SI{2.98}{\mol\per\liter}, which is not small; in fact, it is similar in magnitude to typical reactant concentrations (\emph{c.f.} Ref. \citenum{thomas2016ground}). Solvent concentrations can be an order of magnitude larger and thus come with larger coupling strengths, which may prove more of a challenge to a perturbative approach. However, the importance of intermolecular interactions in a liquid means that one needs to study a large ensemble of solvent molecules to obtain representative results; this means that simulating a solvent fully \emph{ab initio}, either by CBO or by cCBO methods, is completely out of question to start with. A more sensible approach may be to build a model starting from a force field as done in Ref. \cite{li2021collective,lieberherr2023vibrational}, and improve upon it by adding corrections related to the dipole moment fluctuations $\langle\hat{\mu}^2\rangle-\langle\hat{\mu}\rangle^2$ and the polarizability $\alpha$, or other terms based on the cCBO Hamiltonian and perturbative corrections. 

In most of this work, we have assumed there to be only one cavity mode, but in reality typical microcavities support many modes, and since all modes and polarizations are coupled by the perturbative terms (as shown in the SI, and also observed in the CBO approximation \cite{bonini2022ab}), one should actually work in a multi-mode picture. This we leave for future work. 

Finally, one might ask what the future holds for CBO-based approaches. On the one hand, the CBO-Hamiltonian is guaranteed to give the correct result for any coupling strength within the CBO approximation (and the finite-basis approximation at a given level of electronic structure theory).
On the other hand, the cCBO-Hamiltonian and the perturbative framework that comes with it has many advantages over its CBO counterpart. For one, it is straightforward to implement using already available, highly optimized electronic structure packages, its results can be converged by including perturbative corrections and they are in good agreement with CBO results. Additionally, in calculations using the cCBO-Hamiltonian one can `mix-and-match': one could for example take the out-of-cavity frequency from experiment, calculate the ground-state dipole moment at a high level of theory, and the dipole moment squared or transition dipole moments at a lower level of theory; this flexibility is not present in the CBO-framework.
Lastly, and perhaps most importantly, the cCBO-Hamiltonian and its perturbative corrections provide physical insight in the way VSC affects molecules, in terms of what effects are at play, and in what regimes (large or small $N$ and $\eta$) these effects are important. Of course, one can combine both approaches and dissect CBO results in terms of cCBO and perturbative contributions, each of which can possibly be assigned a physical meaning. We hope that this framework can in the future enable a more concrete discussion about what effect the cavity exactly has on molecules.

\section*{Methods}

\subsection{Computational details}
With the MOLPRO 2023 \cite{molpro1,molpro2,molpro3} electronic structure program, we performed Hartree--Fock calculations on HF using an aug-cc-pVDZ basis set, as in Refs. \cite{schnappinger2023PES,schnappinger2023vibr}. To calculate the expectation value $\bra{0}\hat{\mu}^2\ket{0}$, we first write out $\hat{\mu}^2$, giving
\begin{equation}
\begin{aligned}
        \hat{\mu}^2=&\Big(\sum_{i=1}^{N_\text{e}}\hat{r}_i\Big)^2 - 2 \sum_{i=1}^{N_\text{e}}\hat{r}_i\sum_{j=1}^{N_\text{N}} Z_j \hat{R}_j + \Big(\sum_{j=1}^{N_\text{N}} Z_j \hat{R}_j\Big)^2\\
        =&\Big(\sum_{i=1}^{N_\text{e}}\hat{r}_i\Big)^2 + 2 \hat{\mu}_\text{e} \mu_\text{N} + \mu_\text{N}^2\\
\end{aligned}
\end{equation}
where $\hat{\mu}_\text{e}$ and $\mu_\text{N}$ denote the electronic and nuclear dipole moment, respectively. The only non-standard term in this expression is the first term. We can divide it up into one-electron and two-electron contributions:
\begin{equation}
    \Big(\sum_{i=1}^{N_\text{e}}\hat{r}_i\Big)^2=\sum_{i}\hat{r}_i^2 + \sum_{i\neq j}\hat{r}_i \hat{r}_j
\end{equation}
The first term in this expression, which is (a component of) the electronic quadrupole moment, is a property implemented in most electronic structure packages. The second term however is not. For single-determinant wavefunctions, we can evaluate its expectation value in terms of occupied orbitals using the Slater--Condon rules. After performing the sum over spins, we obtain the following expressions in terms of occupied spatial molecular orbitals ${\ket{\phi_n}}$:
\begin{equation}
\begin{aligned}
        \bra{0}\sum_{i\neq j}\hat{r}_i \hat{r}_j\ket{0} = \sum_{n\neq m}^{\text{occupied}} &\Big[4 \bra{\phi_n} \hat{r} \ket{\phi_n} \bra{\phi_m} \hat{r} \ket{\phi_m} \\
        &- 2 \bra{\phi_n} \hat{r} \ket{\phi_m}\bra{\phi_m} \hat{r} \ket{\phi_n}\Big]
\end{aligned}
\end{equation}
To calculate this, we need entries of the matrix $\bra{\phi_n} \hat{r} \ket{\phi_m}$, are anyway computed by electronic-structure programs. As an example, we include a molpro input file that prints these in the SI.

For the evaluation of the PT($\eta^2$) and PT($\eta^3$) corrections, we need transition dipole and $\hat{\mu}^2$ moments. For single-determinant wavefunctions, these are again straightforward to evaluate using the Slater--Condon rules. A first simplification follows from the fact that for one-electron operators (such as $\hat{\mu}$), the only non-vanishing transitions from the ground state are those to singly-excited states. Denoting $\ket{\Psi_m^n}$ to be a state in which a single electron is promoted from orbital $m$ to orbital $n$, we can write
\begin{equation}
    \begin{aligned}
        \bra{0} \hat{\mu}\ket{\Psi_m^n} &= \bra{\phi_m} \hat{r} \ket{\phi_n}\\
        \bra{0} \sum_i \hat{r}_i^2\ket{\Psi_m^n} &= \bra{\phi_m} \hat{r}^2 \ket{\phi_n}\\
        \bra{0} \sum_{i\neq j} \hat{r}_i \hat{r}_j \ket{\Psi_m^n} &= 2 \sum_p\Big[  4\bra{\phi_m} \hat{r} \ket{\phi_n} \bra{\phi_p}\hat{r}\ket{\phi_p}\\
        &\hspace{1cm} - 2 \bra{\phi_m} \hat{r} \ket{\phi_p} \bra{\phi_p}\hat{r}\ket{\phi_n}\Big].\\
    \end{aligned}
\end{equation}
The only additional information we need to calculate this are the $\hat{r}^2$ matrix elements, which can be printed from molpro in much the same way as the $\hat{r}$ matrix elements. An example of this is also included in the molpro input file in the SI. 

Summing over all possible excited states amounts to a double sum over all occupied and all virtual states. For larger molecules, it may be beneficial to truncate the sum, as excitations from core occupied orbitals and to high-lying virtual states come with a larger energy gap, and therefore their contributions to Eq. \ref{eq:PTexpression} are suppressed. 

In the above, we have only considered the dipole moment along a single polarization direction. It is however straightforward to generalize to arbitrary polarization, when noting that $(\hat{\mathbf{e}}_{\lambda} \cdot \hat{\boldsymbol{\mu}})^2=\hat{\mathbf{e}}_{\lambda}^\text{T} (\hat{\boldsymbol{\mu}} \otimes \hat{\boldsymbol{\mu}}) \hat{\mathbf{e}}_{\lambda}$, meaning all information needed is contained in the $3\times 3$ outer-product matrix:
\begin{equation}
    (\hat{\boldsymbol{\mu}} \otimes \hat{\boldsymbol{\mu}}) = \begin{pmatrix}
        \hat{\mu}_x\hat{\mu}_x & \hat{\mu}_x\hat{\mu}_y & \hat{\mu}_x\hat{\mu}_z \\
        \hat{\mu}_y\hat{\mu}_x & \hat{\mu}_y\hat{\mu}_y & \hat{\mu}_y\hat{\mu}_z\\
        \hat{\mu}_z\hat{\mu}_x & \hat{\mu}_z\hat{\mu}_y & \hat{\mu}_z\hat{\mu}_z
    \end{pmatrix}.
\end{equation}

\subsection{Hessians}
Using the indices $i$ and $j$ to label molecular modes, the Hessian for a single cavity mode and $N$ molecules at the minimum geometry with $n_\text{modes}$ vibrational modes each is made up of the following components:
\begin{equation}\label{eq:cCBOhess}
    \begin{aligned} 
    {A}_{ij} &\equiv    \frac{\partial^2 V_\text{BO,0}}{\partial R_i \partial R_j} +  \sqrt{2/\hbar}\, \omega_\mathrm{c}^{3/2} \, q\, \eta\, \frac{\partial^2 }{\partial R_i \partial R_j}\bra{0} \hat{{\mu}}_1\ket{0}  \\
    &+\frac{\omega_\mathrm{c}}{\hbar} \eta^2 \Big[\frac{\partial^2 }{\partial R_i \partial R_j}\bra{0}\hat{{\mu}}_1^2 \ket{0} + \sum_{m\neq 1 }\bra{0} \hat{{\mu}}_m\ket{0} \frac{\partial^2 }{\partial R_i\partial R_j} \bra{0} \hat{{\mu}}_1\ket{0} \Big]\\
    &=  \frac{\partial^2 V_\text{BO,0}}{\partial R_i \partial R_j} +\frac{\omega_\mathrm{c}}{\hbar} \eta^2 \Big[\frac{\partial^2 }{\partial R_i \partial R_j}\bra{0}\hat{{\mu}}_1^2 \ket{0}\\
    &- (N+1) \bra{0} \hat{{\mu}}_1\ket{0} \frac{\partial^2 }{\partial R_i\partial R_j} \bra{0} \hat{{\mu}}_1\ket{0} \Big]\\ 
    {G}_{ij}&\equiv \frac{\omega_\mathrm{c}}{\hbar}\eta^2\frac{\partial }{\partial R_i} \bra{0} \hat{{\mu}}_1\ket{0}\frac{\partial}{\partial R_j} \bra{0} \hat{{\mu}}_1\ket{0} \hspace{2cm} \\
    {B}_i &\equiv \frac{\partial^2 V_\text{cCBO}}{ \partial q \partial R_i} =\sqrt{2/\hbar} \omega_\mathrm{c}^{3/2} \eta \frac{\partial}{\partial R_i}\bra{0}\hat{{\mu}}_1\ket{0} \\
    C &\equiv \frac{\partial^2 V_\text{cCBO}}{ \partial q^2}= \omega_\mathrm{c}^2 .  \\
    \end{aligned}
\end{equation}
Here we have assumed that the molecules are independent, so that the derivative of the dipole moment of one molecule with respect to the coordinates of another molecule vanishes. Also, the operator $\hat{\mu}_1$ denotes a single-molecule dipole moment (in contrast to the ensemble dipole moment $\hat{\mu}$). The derivatives of $\bra{0}\hat{{\mu}}_1\ket{0}$ and  $\bra{0}\hat{{\mu}}_1^2\ket{0}$ are calculated using finite differences. 
Note that $\textbf{A}$ and $\textbf{G}$ are matrices of size $(n_\text{modes}\times n_\text{modes})$, $\textbf{B}$ is a vector of length $n_\text{modes}$ and C is a scalar. Using this, the $N$-molecule Hessian becomes:
\begin{equation}
    \mathbf{H} = \begin{pmatrix}
        \mathbf{A} & \mathbf{G} & \mathbf{G} & \cdots & \mathbf{B} \\
        \mathbf{G} & \mathbf{A} & \mathbf{G} & \cdots & \mathbf{B} \\
        \mathbf{G} & \mathbf{G} & \mathbf{A} & \cdots & \mathbf{B} \\
        \vdots & \vdots & \vdots & \ddots & \vdots \\
        \mathbf{B}^\text{T} & \mathbf{B}^\text{T} & \mathbf{B}^\text{T} & \cdots & C \\
    \end{pmatrix}
\end{equation}
Frequencies can be obtained by mass-weighting this Hessian and then finding the eigenvalues. 

For a single HF molecule, this reduces to a $2\times 2$ matrix, so that we can identify the shifted molecular frequency
\begin{equation}
    \omega_\text{m}'^2 = \omega_\text{m}^2 +\frac{\omega_\mathrm{c}}{\hbar M_\text{HF}} \eta^2 \Big[\frac{\partial^2 }{\partial R^2}\bra{0}\hat{{\mu}}^2 \ket{0}
    - 2 \bra{0} \hat{{\mu}}_1\ket{0} \frac{\partial^2 }{\partial R^2} \bra{0} \hat{{\mu}}_1\ket{0} \Big],
\end{equation}
where $M_\text{HF}$ is the reduced mass.

The intensities can be obtained from the eigenvectors $\mathbf{v}$ of the mass-weighted Hessian. Using that $\partial \expval{\hat{\mu}}/\partial q=0$, the intensity is proportional to
\begin{equation}
    \bigg(\Big(\frac{\partial \expval{\hat{\mu}}}{\partial \mathbf{R}} , \frac{\partial \expval{\hat{\mu}}}{\partial \mathbf{R}} , \frac{\partial \expval{\hat{\mu}}}{\partial \mathbf{R}} , \dots , 0  \Big)^\text{T} \cdot \mathbf{v} \bigg)^2
\end{equation}
As in Ref. \cite{schnappinger2023vibr}, the spectra shown are generated by broadening the signal with a Lorentzian function with a full width at half maximum of \SI{10}{\per\centi\meter}. The signals are normalized such that the intensities of the two peaks add up to 2. 

The PT($\eta^2$) correction to the cCBO Hamiltonian is conveniently written as
\begin{equation}
\begin{aligned}
        E_0^{(2),\eta^2}&=\frac{2 \omega_\mathrm{c}^2 \eta^2}{\hbar} q^2 \sum_{n\neq 0 } \frac{\,\bra{0}\hat{\mu}\ket{n}^2}{V_0(\mathbf{R}) - V_n(\mathbf{R}) } \\
        &\equiv -\frac{ \omega_\mathrm{c}^2 \eta^2}{\hbar} q^2  \alpha_1(\mathbf{R})
\end{aligned}
\end{equation}
for a single molecule. Note that the polarizability $\alpha_1(\mathbf{R})$ is always positive, meaning that this correction is always negative. For $N$ molecules, the dipole operator is just the sum of all single-molecule dipole operators, and therefore only transition dipole moments to single-molecule excited states are non-zero; this means we can write $\alpha_N(\mathbf{R}_1, \mathbf{R}_2, ...) = \sum_i \alpha_1(\mathbf{R}_i)$. Using this, we find that the PT($\eta^2$) correction adds the following terms to Eq. \ref{eq:cCBOhess}:
\begin{equation}
        \begin{aligned}
A_{ij}^\text{PT}&= -\frac{\eta^2\omega_\mathrm{c}^3}{\hbar}q^2  \frac{\partial^2}{\partial R_i \partial R_j}\alpha_1(\textbf{R})\\
&=-\eta^4\frac{\omega_\mathrm{c}^2}{\hbar^2}N^2\bra{0}\hat{\mu}_1\ket{0}^2\frac{\partial^2}{\partial R_i \partial R_j}\alpha_1(\textbf{R}) \\
B_i^\text{PT}
&= -\frac{2\eta^2\omega_\mathrm{c}^3}{\hbar}q \frac{\partial}{\partial R_i} \alpha_1(\textbf{R})\\
&=\eta^3 \sqrt{\frac{8\omega_\text{c}^5}{\hbar^3}} N \bra{0}\hat{\mu}_1\ket{0}\frac{\partial}{\partial R_i} \alpha_1(\textbf{R})\\
C^\text{PT} &= -\frac{2\eta^2\omega_\mathrm{c}^3}{\hbar}\alpha_N(\textbf{R}) \\ &= -\eta^2\frac{2\omega_\mathrm{c}^3}{\hbar}N\alpha_1(\textbf{R}).
\end{aligned}
\end{equation}
The derivatives of $\alpha_1(\mathbf{R})$ are calculated with finite differences.

The Hessian of the $\langle \hat{\mu} \rangle^2$-Hamiltonian is the same as that for the cCBO Hamiltonian, except for a small change in the matrix $\mathbf{A}$ in Eq. \ref{eq:cCBOhess}. It now reads:
\begin{equation}\label{eq:mu2hess}
    \begin{aligned} 
    {A}^{\langle\hat{\mu}\rangle^2}_{ij} &\equiv    \frac{\partial^2 V_\text{BO,0}}{\partial R_i \partial R_j} +  \sqrt{2/\hbar}\, \omega_\mathrm{c}^{3/2} \, q\, \eta\, \frac{\partial^2 }{\partial R_i \partial R_j}\bra{0} \hat{{\mu}}_1\ket{0}  \\
    &+2\frac{\omega_\mathrm{c}}{\hbar} \eta^2 \Big[\frac{\partial\bra{0}\hat{\mu}_1\ket{0} }{\partial R_i}\frac{\partial \bra{0}\hat{\mu}_1\ket{0}}{\partial R_j} \\
    &+ N\bra{0}\hat{\mu}_1\ket{0}\frac{\partial^2 }{\partial R_i \partial R_j} \bra{0}\hat{\mu}_1\ket{0} \Big]\\
    & = \frac{\partial^2 V_\text{BO,0}}{\partial R_i \partial R_j} +2\frac{\omega_\mathrm{c}}{\hbar} \eta^2 \frac{\partial\bra{0}\hat{\mu}_1\ket{0} }{\partial R_i}\frac{\partial \bra{0}\hat{\mu}_1\ket{0}}{\partial R_j} \\
\end{aligned}
\end{equation}

The generalization to many cavity modes and polarizations straightforward; then $\textbf{B}$ and $\textbf{C}$ will also become matrices. 

For $N$ molecules, we can use the structured nature of the Hessian in order to block diagonalize it. Only the symmetric linear combination of molecules, which we denote as $ (\mathbf{\mathsf{1}} \, \mathbf{\mathsf{1}}\, \mathbf{\mathsf{1}}\, ... \,\mathbf{\mathsf{1}})/\sqrt{N}$ with $\mathbf{\mathsf{1}}$ the $n_\text{modes} \times n_\text{modes}$ identity matrix, will couple to the cavity, with a coupling strength of $N\mathbf{B}/\sqrt{N} = \sqrt{N}\mathbf{B}$. Its matter--matter block will be given by $(N\mathbf{A}+N(N-1)\mathbf{G})/N=\mathbf{A}+(N-1)\mathbf{G}$.  All other linear combinations decouple, meaning the relevant part of the Hessian is
\begin{equation}
    \mathbf{H} = \begin{pmatrix}
        \mathbf{A} +(N-1)\mathbf{G} & \sqrt{N}\,\mathbf{B} \\
        \sqrt{N}\,\mathbf{B}^\text{T}  & C \\
    \end{pmatrix},
\end{equation}
which is easily diagonalized numerically.

\subsection{A note on the comparison of spectra}
When comparing to the spectral data of Ref. \cite{schnappinger2023vibr}, we noticed that their cavityless harmonic frequency is about \SI{0.4}{\per\centi\meter} higher than ours (their frequency being \SI{4467.22}{\per\centi\meter}, ours \SI{4466.84}{\per\centi\meter}. To make up for this, we moved their results down in frequency by \SI{0.4}{\per\centi\meter}, and shifted our cavity frequency down by \SI{0.4}{\per\centi\meter} to \SI{4466.6}{\per\centi\meter} (rather than exactly \SI{4467.0}{\per\centi\meter}). This small shift in cavity frequency has a subtle effect on the peak heights.

\section*{Acknowledgements}
The authors would like to thank Thomas Schnappinger and Markus Kowalewski for kindly sharing their data with us. M.R.F. was supported by an ETH Zurich Research Grant.


\section*{Competing interests}
The authors declare no competing interests.
\clearpage



\section*{Supplementary material}

# Supplementary Information: Understanding the Cavity Born–Oppenheimer Approximation

Marit R. Fiechter and Jeremy O. Richardson

January 7, 2024

## 1  Second-order perturbation theory for more than one cavity mode

In the main paper, we have given the energy correction arising from second order perturbation theory only for a single cavity mode and polarization direction. For an arbitrary number of cavity modes (with index $k$) and two polarization directions (index $\lambda$) per mode, we can express the result more generally as

$$
\begin{aligned}
E_0^{(2)} &= \sum_{n\neq 0} \frac{\left( \sum_{\lambda,k} \left[ \sqrt{2/\hbar}\,\omega_{\mathrm{c}}^{3/2}\eta\, q_{\lambda,k}\, k\, \hat{\mathbf{e}}_\lambda \cdot \langle 0|\,\hat{\boldsymbol{\mu}}\,|n\rangle + \frac{\omega_{\mathrm{c}}}{\hbar}\eta^2 \langle 0|\,(\hat{\mathbf{e}}_\lambda \cdot \hat{\boldsymbol{\mu}})^2\,|n\rangle \right] \right)^2}{V_0(\mathbf{R}) - V_n(\mathbf{R}) + \eta \sum_{\lambda,k} \left[ \sqrt{2/\hbar}\,\omega_{\mathrm{c}}^{3/2}\, k\, q_{\lambda,k}\, \hat{\mathbf{e}}_\lambda \cdot (\langle 0|\,\hat{\boldsymbol{\mu}}\,|0\rangle - \langle n|\,\hat{\boldsymbol{\mu}}\,|n\rangle) \right]} + O(\eta^2) \\
&= \sum_{n\neq 0} \eta^2\, \frac{2\omega_{\mathrm{c}}^3}{\hbar} \frac{\left( \sum_{\lambda,k} k\, q_{\lambda,k}\, \hat{\mathbf{e}}_\lambda \cdot \langle 0|\,\hat{\boldsymbol{\mu}}\,|n\rangle \right)^2}{V_0(\mathbf{R}) - V_n(\mathbf{R})} \\
&\quad + \eta^3\, 2\sqrt{\frac{2\omega_{\mathrm{c}}^5}{\hbar^3}} \frac{\left[ \sum_{\lambda,k} \langle 0|\,(\hat{\mathbf{e}}_\lambda \cdot \hat{\boldsymbol{\mu}})^2\,|n\rangle \right] \left[ \sum_{\lambda',k'} k'\, q_{\lambda',k'}\, \hat{\mathbf{e}}_{\lambda'} \cdot \langle 0|\,\hat{\boldsymbol{\mu}}\,|n\rangle \right]}{V_0(\mathbf{R}) - V_n(\mathbf{R})} \\
&\quad + \eta^4\, \frac{\omega_{\mathrm{c}}^2}{\hbar^2} \frac{\left( \sum_{\lambda,k} \langle 0|\,(\hat{\mathbf{e}}_\lambda \cdot \hat{\boldsymbol{\mu}})^2\,|n\rangle \right)^2}{V_0(\mathbf{R}) - V_n(\mathbf{R})} \\
&\quad - \eta^3 \sqrt{\frac{8\omega_{\mathrm{c}}^9}{\hbar^3}} \frac{\left[ \sum_{\lambda,k} k\, q_{\lambda,k}\, \hat{\mathbf{e}}_\lambda \cdot (\langle 0|\,\hat{\boldsymbol{\mu}}\,|0\rangle - \langle n|\,\hat{\boldsymbol{\mu}}\,|n\rangle) \right] \left( \sum_{\lambda',k'} k'\, q_{\lambda',k'}\, \hat{\mathbf{e}}_{\lambda'} \cdot \langle 0|\,\hat{\boldsymbol{\mu}}\,|n\rangle \right)^2}{(V_0(\mathbf{R}) - V_n(\mathbf{R}))^2} \\
&\quad + O(\eta^4)
\end{aligned}
\tag{S1}
$$

Here, we again assumed all transition moments to be real, as that is true for transitions to bound states and it allows us to separate the results into terms by orders of $\eta$, thereby providing us with more insight.

## 2  Example of a Molpro input file

In the main text, we mentioned that we obtain the quantities $\langle \phi_n|\,\hat{r}\,|\phi_m\rangle$ and $\langle \phi_n|\,\hat{r}^2\,|\phi_m\rangle$ required for our calculations directly from Molpro 2023. Here we provide an input file that prints (the $x$-component of) these quantities, as well as the orbital energies, to high precision. Note that in Molpro, the $\sum_i \hat{r}_i^2$ operator (or rather, $-\sum_i \hat{r}_i^2$) is referred to as the second moment.

```
    ***,hf properties
print,basis
symmetry,nosymmetry,noorient
angstrom
geometry={
2

H 0                0 0
F 0.900183796749 0 0
}
basis=aug-cc-pVDZ
hf

{matrop
! load orbital coefficients and basis function transition moments
load,orbcoef,orb
load,dipmomx,oper,dmx
load,secmomx,oper,xx
! transform to MO basis and write to file
tran,dipmomxMO,dipmomx,orbcoef
tran,secmomxMO,secmomx,orbcoef
write,dipmomxMO,filename=dipmomxMO.out,status=new,format=float
write,secmomxMO,filename=secmomxMO.out,status=new,format=float
! Recalculate the orbital energies and print to file
load,SMH
load,DEN
fock,F,DEN
tran,Fp,F,SMH
Diag,eigvecs,eigvals,Fp
write,eigvals,filename=orbens.out,status=new,format=float
}


! printing some ground-state properties to the output file
! this is convenient for consistency checks
{property
orbital
density
dm
sm
}
```

# 3   Changes in spectra with increasing the number of molecules $N$

Figs. S1–S5 show for each method how the harmonic spectra change upon increasing the number of molecules (while rescaling the coupling strength according to $\eta = \eta_1/\sqrt{N}$, as discussed in the main text). For completeness, we also show the behaviour of the CBO harmonic spectrum from Ref. [1] here.

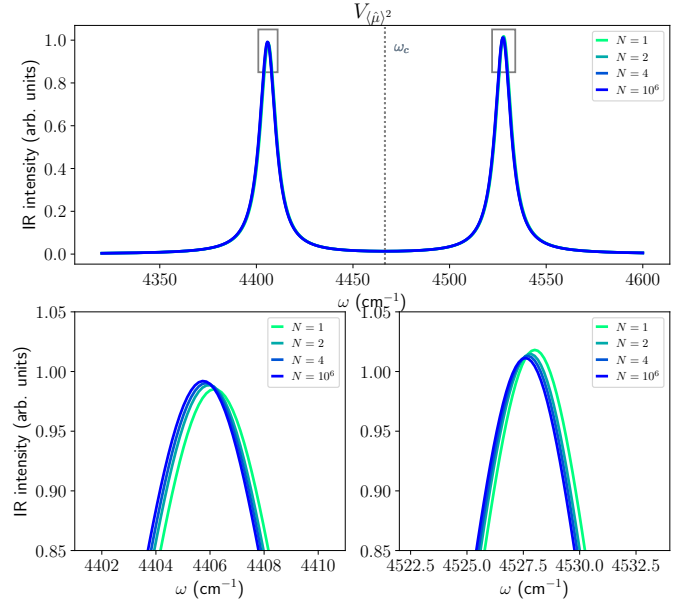

**Figure S1:** Harmonic spectra for the $V_{\langle\hat{\mu}\rangle^2}$ potential at a coupling strength of $\epsilon_c = 3.0$ V/nm.

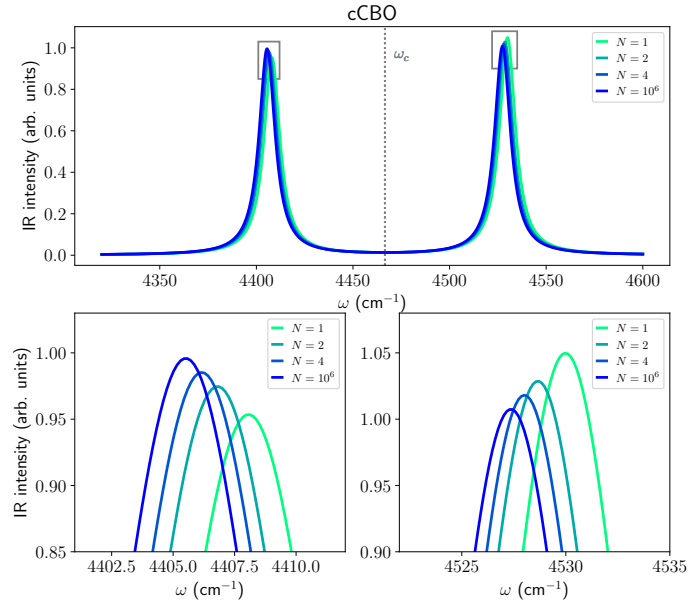

**Figure S2:** Harmonic spectra for the cCBO potential at a coupling strength of $\epsilon_c = 3.0$ V/nm.

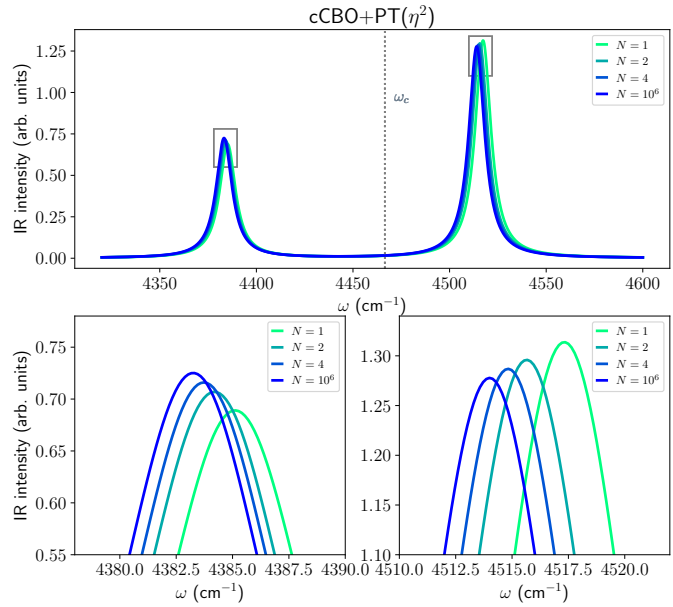

**Figure S3:** Harmonic spectra for the cCBO potential with perturbative corrections up to $\eta^2$, at a coupling strength of $\epsilon_c = 3.0$ V/nm.

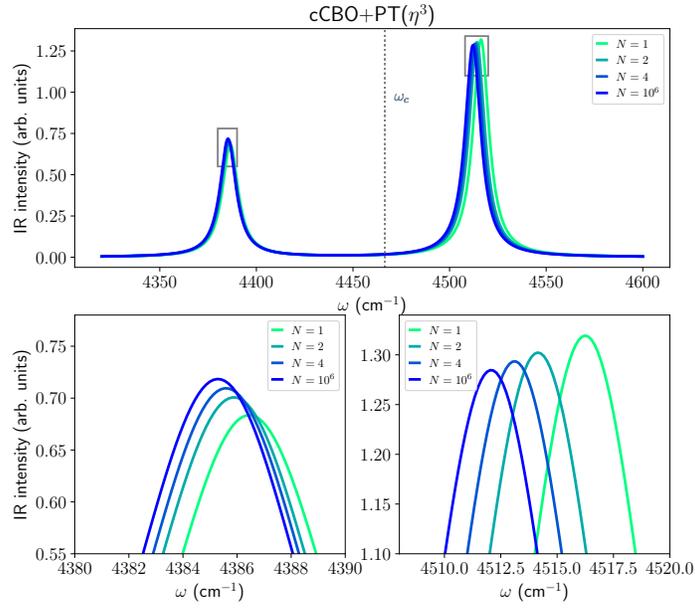

**Figure S4:** Harmonic spectra for the cCBO potential with perturbative corrections up to $\eta^3$, at a coupling strength of $\epsilon_c = 3.0$ V/nm.

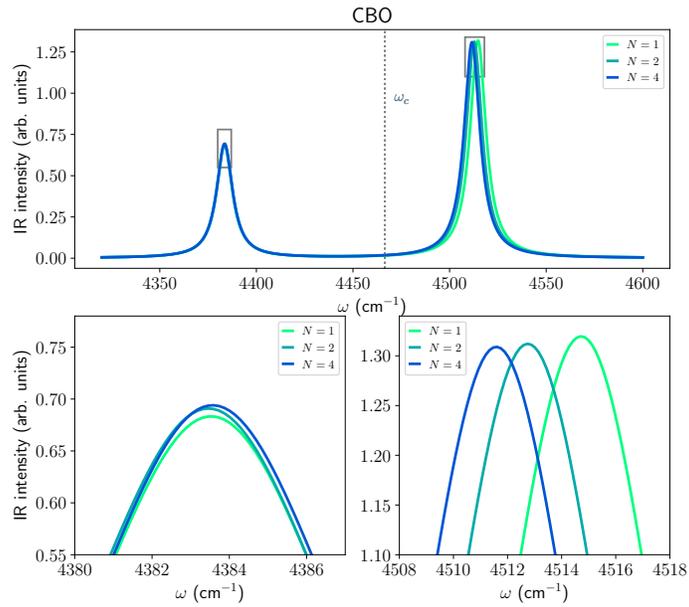

**Figure S5:** Harmonic spectra for the CBO potential at a coupling strength of $\epsilon_c = 3.0$ V/nm; data from Ref. [1].



\end{document}